\documentclass[12pt,oneside]{article}
\usepackage{amsfonts,amssymb,graphicx}
\def\be{\begin{equation}}
\def\ee{\end{equation}}
\def\bea{\begin{eqnarray}}
\def\eea{\end{eqnarray}}
\def\<{\langle}
\def\>{\rangle}
\def\~{\tilde}
\def\s{\sigma}

\def\L{\Lambda}

\def\b{\beta}

\newcommand{\Z}{\Bbb Z}

\newtheorem{theorem}{Theorem}

\begin{document}
\begin{center}
{\bf\sc\Large
surface terms on the nishimori line of the \\ 
gaussian edwards-anderson model}

\vspace{.5cm}

{Pierluigi Contucci$^\dag$,\, Satoshi Morita$^\ddag$,\, 
Hidetoshi Nishimori$^\ddag$}

\vspace{.5cm}

{\small $^\dag$ Dipartimento di Matematica, Universit\`a di Bologna, Italy}

{\small {e-mail: contucci@dm.unibo.it}}

{\small $^\ddag$ Department of Physics, Tokyo Institute of Technology, Japan}
    
{\small {e-mail: smorita@stat.phys.titech.ac.jp}}

\vskip 1truecm
\date{}

\end{center}

\vskip 1truecm
\begin{abstract}\noindent
For the Edwards-Anderson model we find an integral representation for
some surface terms on the Nishimori line. Among the results are
expressions for the surface pressure for free and periodic boundary
conditions and the adjacency pressure, i.e., the difference between the
pressure of a box and the sum of the pressures of adjacent sub-boxes in
which the box can been decomposed.  We show that all those terms indeed
behave proportionally to the surface size and prove the existence in
the thermodynamic limit of the adjacency pressure.
\end{abstract}
\newpage
\section{Introduction}
The next to the leading term in the volume for typical extensive
quantities in statistical mechanics like the free energy is usually
expected to behave as a surface at least for finite-dimensional (non
mean-field) models and regular potentials. This fact has been analyzed
since the seminal work by Fisher and Lebowitz \cite{FL} on classical
particle systems and followed by a series of results in Euclidean
quantum field theories \cite{G,GRS} and in ferromagnetic spin models
\cite{FC}.

Much less is known in the spin glass case. In the Edwards-Anderson model
for instance the proof that the correction size behavior is indeed a
surface came only very recently \cite{CG1,CG2} stemming from the
spectacular interpolation technique which changed the landscape of the
rigorous results in the mean field spin glass models \cite{GT,G2}.

In this paper we perform the analysis of three surface terms for a
finite-dimensional model with a non-centered quenched interaction. Our
study is made within a generalization of the Nishimori line (see definition in Sect. 2), 
in which several exact results can be obtained from the gauge
symmetries of the model \cite{N,N2} (see also the appendix where we
provide a self-contained treatment of some aspects). The first term we
consider is the adjacency pressure, i.e., the difference, for free boundary
conditions, between the pressure of a box and the sum of the pressures
of the disjoint adjacent sub-boxes in which the large box has been
decomposed (see \cite{CF} for its treatment in the ferromagnetic case).
We show in Theorem 1 that such a term behaves indeed like the total
surface of contact between the sub-boxes and that its value per unit
surface exists in the thermodynamic limit. Our method produces in fact
an integral representation for its limiting value and is strongly based
on an inequality which has been recently proved \cite{MNC} to hold on
the NL and which allows to reach in this case the same level of control
which one obtains in ferromagnetic models where the second Griffiths
inequality \cite{Gr} holds true. To our knowledge this is the only case
in disordered systems in which a surface quantity per site can be shown to
exist. We consider also two other surface terms: the difference of the
pressure between free and periodic boundary conditions and the pure
surface pressure on free boundary conditions.  Our method (see Theorems
2 and 3) leads in this case to a control on the size of these terms which 
turns out to be indeed that of the surface of the box.

\section{Surface Terms}
For Ising spins
$S_n$, $n\in \L \subset\Z^d$,  we consider a nearest neighbor potential
\be
 U \ = \ \sum_{b\in B(\L)}\beta_b J_b S_b \;  ,
\label{nnp}
\ee
where $B(\L)$ is the set of bonds $b=(n,n')$, i.e., the couples of nearest
neighbor sites $n$ and $n'$ in $\L$, the parameter $\beta_b\ge 0$ are
local inverse temperatures, $S_b=S_nS_{n'}$, and the $J_b$ are independent
Gaussian random variables defined by the average
\be
 [J_b] \ = \ \mu_{b} \ge 0 \; , 
\ee
and the variance
\be
 [(J_b-\mu_b)^2] \ = \ \s^2_b \; .
 \label{variance}
\ee
The generalized NL for the present case of local temperature is defined by the two
conditions $\b_b=x_b/\s_b$ and $\mu_b=\s_b x_b$ for every choice of
$x_b\ge 0$ \cite{MNC}.

By definition the quenched pressure of the model is defined as
\be
P \ = \ \int\prod_b \frac{dJ_b}{\sqrt{2\pi \s_b^2}}
e^{-\frac{(J_b-\mu_b)^2}{2\s^2_b}}
\ln\sum_{\{S\}}e^{\sum_{b}\beta_b J_b S_b} \; ;
\ee 
on the NL one can introduce the Gaussian variable $j_b=J_b/\s_b$ whose
mean is $x_b$ and variance is $1$ and observe that the pressure admits
the representation
\be
P(\{x\}) \ = \ \int\prod_b \frac{dj_b}{\sqrt{2\pi}}e^{-\frac{(j_b-x_b)^2}{2}}
\ln\sum_{\{S\}}e^{\sum_{b}x_b j_b S_b} \; ,
\ee
where the notation is stressing the fact that on the NL the only parameters
the pressure depends on are the positive numbers $x_b$.

The first surface term we want to consider is the difference, at inverse
temperature $\beta$, between the pressure of a box $\L$ and the sum of
the pressures for a disjoint decomposition of it:
$\L=\bigcup_{i}\Lambda_i$, where $\Lambda$ and $\Lambda_i$ are both with
free boundaries. We expect such a quantity to be of the size of the
corridor $\cal C$, i.e., the set of bonds joining neighboring regions of
the $\L_i$ and we will call it $T_{\cal C}$. To prove it we choose the
family $\{x\}$ according to the law
\be
x_b \; = \; \left\{
\begin{array}{ll} 
\b\sigma\sqrt{t}, & \mbox{if $b\in\cal C$}, \\
 \b\sigma, & \mbox{otherwise} \; .
\end{array}\right. 
\label{cut}
\ee
With this choice the pressure is a function of $\b$ and $t$,
$P(\b,t)$. Clearly $P(\b,1)=P_\L(\b)$, where $P_\L(\b)$ is the pressure
of a box $\L$ at the inverse temperature $\b$ on the NL and analogously
$P(\b,0)=\sum_i P_{\L_i}(\b)$ since the zero value of the interaction
along the corridor makes the subsets independent.  By the fundamental
theorem of calculus
\be
T_{\cal C} \; = \; P_\L(\b)-\sum_i P_{\L_i}(\b) 
\; = \; P(\b,1) -P(\b,0) 
\; = \; \int_{0}^{1} \frac{\partial}{\partial t}P(\b,t) dt \; .
\label{tici}
\ee
Moreover 
\be
\frac{\partial}{\partial t}P(\beta, t) \, = \, 
\sum_b \frac{\partial P(\{x\})}{\partial x_b}\frac{dx_b}{dt} \; ,
\label{dfc}
\ee
where $x_b$ is set according to (\ref{cut}) after differential
${\partial P}/{\partial x_b}$. From $(\ref{cut})$ we obtain
\be
\frac{dx_b}{dt} \; = \; \left\{
\begin{array}{ll} 
\b\sigma/2\sqrt{t}, & \mbox{if $b\in\cal C$}, \\
0, & \mbox{otherwise} \; .
\end{array}\right. 
\label{ddix}
\ee
Introducing the Boltzmann-Gibbs expectation at fixed disorder
\be
\<S_b\> \, = \, \frac{\sum_{\{S\}} S_b e^U}{\sum_{\{S\}} e^U},
\label{bg}
\ee
we can show by a straightforward computation (see appendix) that 
\be
\frac{\partial P}{\partial x_b} \; = \; x_b [\<S_b+1\>] \; .
\label{a1}
\ee
Putting
together $(\ref{tici}),(\ref{dfc}),(\ref{ddix})$ and $(\ref{a1})$ we obtain
\be
T_{\cal C} \; = \; \frac{\b^2\sigma^2}{2}\sum_{b\in {\cal C}}
\int_{0}^{1} [\<S_b + 1\>_t]_t dt \; ,
\ee
where we have indicated explicitly the dependence on the parameter $t$
of both the Gaussian integration $[\cdots]$ which depends on $t$ through
its mean $x_b$ and the Boltzmann-Gibbs state $\<\cdots \>$ which depends
on it through the potential. Let us now write $\beta \sigma=x$. Defining
the average bond-spin on the corridor as 
$S_{\cal C}=|{\cal C}|^{-1}\sum_b S_b$ we find
\be
T_{\cal C} \; = \; |{\cal C}|\frac{x^2}{2}
\left( 1 + \int_{0}^{1} [\<S_{\cal C}\>_t]_t dt\right)
\; .
\label{rep1}
\ee
If, for instance, we choose $\L$ to be a $d$-dimensional square box of
side $2L$ and the sets $\L_i$ to be the $2^d$ hypercubes which compose
it, we immediately find that the cardinality of $\cal C$ is equal to the
surface of $\L$, i.e., $|{\cal C}|=d(2L)^{d-1}$ up to a lower-order term
(${\cal O}(L^{d-2})$). In this case we can write with obvious meaning of
the symbols
\be
\frac{T_L}{L^{d-1}} \; = \; d x^2 2^{d-2}\left( 1 + \int_{0}^{1}
[\<S_{\cal C}\>_t]_t dt
\right)
\; .
\label{tau1}
\ee
This representation allows us to show that the limit of the quantity on
the left-hand side when $L$ goes to $\infty$ does exist. In fact it is
possible to see that each correlation $[\<S_b\>]$ is, on the NL, monotonic
increasing with respect to any $x_{b'}$:
\be
\frac{d}{dx_{b'}}[\<S_b\>] \; \ge \; 0 \; .
\label{g2}
\ee
This inequality originally proved in \cite{MNC} (see also the appendix)
plays the same role as the second Griffiths inequality for ferromagnetic
systems \cite{Gr} and leads immediately, by a similar argument,
to the existence of the thermodynamic limit of the correlation functions 
for free boundary conditions. From the existence of the bond correlation functions it is
possible to deduce the existence of the quantity $[\<S_{\cal C}\>_t]_t$,
which turns out to coincide with its thermodynamically relevant component,
i.e., the value of the bond correlation function away from the boundary
of each cube hyperface. The proof runs in full parallel to the
ferromagnetic case and is treated, for instance, in \cite{FC} (see also
\cite{Si} Theorem II.8.3).

The arguments developed so far can be summarized as follows:

\begin{theorem}
%[Existence and integral representation]
\label{main1}
On the NL the term $T_L$ produced by subtracting from the pressure of a
square d-dimensional box of side $2L$ (with free boundaries) those of
the composing sub-boxes of side $L$ grows in proportion to the surface
size $L^{d-1}$. The ratio
\be
\frac{T_L}{L^{d-1}}
\ee
exists in the thermodynamic limit and admits the representation
\be 
dx^2 2^{d-2}\left( 1 + \int_{0}^{1}
[\<S_1S_2\>_t]_t dt
\right)
\ee
where the two spins $S_1$ and $S_2$ are two nearest neighbors of two
adjacent boxes far away from the outer boundaries. The quenched states
are computed for free boundary conditions and with the interpolation
scheme $(\ref{cut})$.
\end{theorem}

It is interesting to observe that while the same term is known to have a
surface size also in the Edwards-Anderson model with symmetric
randomness \cite{CG2} its existence is still an open question due to
the lack of an inequality of the type of the second Griffiths inequality
which would ensure the existence by monotonicity of the correlation
functions.

The next term we want to estimate is the difference between the quenched
pressure for periodic ($\Pi$) and free ($\Phi$) boundary conditions.  In
order to find its size dependence we consider the $d$-dimensional torus of
side $L$, $\Pi_L$, and we define $\cal C$ to be the standard cut of the
torus, i.e., the set of bonds cutting along which the torus unfolds into
the hypercube $\L_L$. The argument proceeds like in the previous case with
formally the same definition of the interpolating parameters $\{x\}$ as
in $(\ref{cut})$. In this case we have
\be
\label{fr_pe}
T^{(\Pi,\Phi)}_{L} \; = \; P_{\Pi_L}(\b)-P_{\L_L}(\b) 
\; = \; P_{\Pi_L}(\b,1) - 
P_{\Pi_L}(\b,0) \; = \; \int_{0}^{1} 
\frac{\partial}{\partial t}P_{\Pi_L}(\b,t) dt \; .
\ee
Since the cardinality of $\cal C$ is now $dL^{d-1}$, $(\ref{rep1})$ lead to
\be
T^{(\Pi,\Phi)}_{L} \; = \; \frac{1}{2}d x^2L^{d-1}
\left( 1 + \int_{0}^{1} [\<S_{\cal C}\>^{(\Pi_L)}_t]_t dt
\right)
\; ,
%\label{rep1}
\ee
where the brackets represent the Boltzmann-Gibbs state at fixed
disorder for periodic boundary conditions and the quenched states are
computed in the suitable interpolation scheme. We can thus formulate the
following theorem:
\begin{theorem}
\label{main2} 
On the NL the term $T^{(\Pi,\Phi)}_{L}$ obtained by subtracting from
the pressure for periodic boundary conditions the one with free boundary
conditions grows like a surface and admits the integral representation
\be 
\frac{1}{2}d x^2L^{d-1}\left( 1 + \int_{0}^{1}
[\<S_{\cal C}\>^{(\Pi_L)}_t]_t dt
\right) \; .
\ee
where the quenched states are computed for periodic boundary conditions
and with the interpolation scheme $(\ref{cut})$ along the standard cut of
the torus.
\end{theorem}
Although the inequality $(\ref{g2})$ holds in complete generality, one
cannot deduce from it the existence by monotonicity of the correlation
functions for periodic boundary conditions and, by consequence, the
existence result of the surface term in the thermodynamic limit is not
within reach at present. One encounters the same difficulty in the
ferromagnetic case (see again \cite{Si} Theorem II.8.3).

The final expression we want to consider represents the genuine {\it
surface pressure} defined, for boundary conditions $\Xi (=\Pi$ or $\Phi$), as (see
\cite{Si})
\be
T^{(\Xi)}_{\partial\L} \; = \; P^{(\Xi)}_\L - p|\Lambda| \; ,
\label{qsp} 
\ee
where $p$ is the limiting pressure per site. The existence by monotonicity 
of the quantity $p$ on the NL has been proved for free boundary conditions in
\cite{MNC}. The result of the previous theorem tell us that the difference 
$P_{\Pi_L}(\b)-P_{\L_L}(\b)$ is of the order of a surface: by consequence the 
pressure per site $p_L=P^{(\Xi)}_\L/L^{d}$ converges to the same value $p$
for both free and periodic boundary conditions. One may prove along the same
lines (using interpolation between two assigned boundary conditions as in
(\ref{cut}) and taking the fact into account that (\ref{a1}) is valid for
any boundaries) that the limiting pressure exists for a wide class of boundary 
conditions and its value is the same for all of them.

Let us take a square box $\L$ of side $L$ and its $k$-magnification,
i.e., the box of side $kL$. For free boundary conditions
\be
P^{(\Phi)}_L \; = \; [\ln Z^{(\Phi)}_L] \; = \; k^{-d}[\ln
(Z^{(\Phi)}_L)^{k^d}]\; .
\label{fbc}
\ee
On the other hand the same limiting pressure can be obtained for
instance for periodic boundary conditions
\be
pL^d \; = \; \lim_{k\to\infty} k^{-d} [\ln Z^{(\Pi)}_{kL}] \;
\label{pbc}
\ee
By (\ref{fbc}) and (\ref{pbc}) we obtain
\bea\nonumber
T^{(\Phi)}_{\partial\Lambda} \, &=&  \left( P^{(\Phi)}_L - pL^d \right) \\
&=& \, \lim_{k\to \infty} k^{-d}\left[\ln (Z^{(\Phi)}_L)^{k^d}- \ln
Z^{(\Pi)}_{kL}\right] \; .
\label{bs}
\eea
In this case we define the corridor $\cal C$ as the set of nearest
neighbors belonging to adjacent boxes of side $L$ which compose the
large torus of side $kL$. Following the same procedure as before we
arrive at
\be
T^{(\Phi)}_{\partial\Lambda} \, = \, 
- \lim_{k\to \infty} k^{-d} \int_{0}^{1}
\frac{\partial}{\partial t}P_{\Pi_{kL}}(\b,t)dt \; ;
\label{bs2}
\ee
in this case the cardinality of $\cal C$ turns out to be $dL^{d-1}k^d$
which, together with the computation of the derivative, leads us to the
following result:
\begin{theorem}
\label{main3}
On the NL the surface pressure for free boundary conditions grows like a
surface and admits the integral representation
\be
T^{(\Phi)}_{\partial \L} \; = \; 
- \frac{d}{2}x^2 L^{d-1}\left( 1 + \int_{0}^{1} 
[\<S_{\cal C}\>^{(\Pi_{\infty L})}_t]_t dt
\right)
\; ,
\label{replast}
\ee
where the brackets represent the Boltzmann-Gibbs state of an infinite
system for periodic boundary conditions made up of boxes of side $L$ and
interpolated along the adjacent bonds. Since 
$T^{(\Pi,\Phi)}_{L} = T^{(\Pi)}_{\partial \L}-T^{(\Phi)}_{\partial \L}$,
we may use Theorem 2 and $(\ref{replast})$ to obtain
\be
T^{(\Pi)}_{\partial \L} \; = \; \frac{1}{2}dx^2L^{d-1}
\left(\int_{0}^{1}
[\<S_{\cal C}\>^{(\Pi_L)}_t]_t dt - \int_{0}^{1}
[\<S_{\cal C}\>^{(\Pi_{\infty L})}_t]_t dt
\right) \; .
\ee
\end{theorem}
It is interesting to observe that on the NL of the Edwards-Anderson
model the surface pressure fulfils the inequalities
$T^{(\Phi)}_{\partial\Lambda}\le 0$ which is the same as in the
ferromagnetic case.  This fact is totally non trivial a priori because
the interactions have no definite sign in our model and the Griffiths
inequalities do not hold in the standard form; nevertheless the
positivity of the average of the interaction is enough to guarantee that
the sign of the surface pressure persists in the disordered case.

\section{Conclusion}
In this paper, we have investigated some surface terms on the NL for the
Edwards-Anderson model: the surface pressure for free and periodic
boundary conditions, the difference of the pressure between these
boundary conditions and the adjacency pressure, which is defined as the
difference between the pressure of the box and the sum of the pressure
for its disjoint decomposition. Each of them has been shown to have an integral
representation which is proportional to the surface size.  Moreover we
showed that the adjacency pressure has a thermodynamic limit.  These
results are similar to the ferromagnetic case, but is
quite non-trivial in disordered systems.

\section{Appendix}

In this appendix, we prove the Griffiths inequalities for the Gaussian
Edwards-Anderson model on the NL following \cite{MNC}:
\begin{equation}
 \frac{dP}{dx_b} \; = \; x_b [\<S_b+1\>] \geq 0\; , \label{G1}
\end{equation}
\begin{equation}
 \frac{d}{dx_{b'}}[\<S_b\>] \; = \; 2x_{b'}
   [(\<S_b S_{b'}\>-\<S_b\>\<S_{b'}\>)^2] \; \geq 0 \; . \label{G2}
\end{equation}

Before the proof of these inequalities, we prove the following identities
which hold on the NL:
\begin{equation}
 [\<j_b S_b\>] = x_b, \label{le}
\end{equation}
\begin{equation}
 [\<S_b\>] = [\<S_b\>^2] . \label{mq}
\end{equation}
To obtain the first identity, let us consider the average of the Gaussian
random variable $j_b$,
\begin{equation}
 [j_b] =  \int\prod_{b'} \frac{dj_{b'}}{\sqrt{2\pi}}
  e^{-\frac{(j_{b'}-x_{b'})^2}{2}} j_b .
\end{equation}
The right-hand side does not change under a local transformation
$j_b\rightarrow j_b \sigma_b$, where $\sigma_b$ (to be distinguished
from  $\sigma_b$ in (\ref{variance})) is a product of two
Ising spins at the edge of the bond $b=(n,n')$, that is,
$\sigma_b=\sigma_n \sigma_{n'}$. Thus it can be rewritten using the nearest
neighbor potential (\ref{nnp}),
\begin{equation}
 [j_b] =  \int\prod_{b'} \frac{dj_{b'}}{\sqrt{2\pi}}
  e^{-\frac{j_{b'}^2+x_{b'}^2}{2}}
 e^U j_b \sigma_b .
\end{equation}
Since the value of $\sigma_n$ is arbitrary, we can sum this
equation over all possible values of $\{\sigma_n\}$ and divide the
result by $2^N$ ($N$ is the number of spins). We then obtain using
the Boltzmann-Gibbs expectation (\ref{bg}),
\begin{equation}
 [j_b] =  \int\prod_{b'} \frac{dj_{b'}}{\sqrt{2\pi}}
  e^{-\frac{j_{b'}^2+x_{b'}^2}{2}}
 \sum_{\{\sigma\}} e^U \<j_b S_b\> ,
\end{equation}
where we have written $S_n$ for $\sigma_n$. The right-hand side of the
above equation is equal to $[\<j_b S_b\>]$. This fact can be confirmed
by applying the local transformation to $[\<j_b S_b\>]$. We note that
the expectation $\<j_b S_b\>$ does not change under this transformation
because the transformation for spins, $S_n \rightarrow S_n \sigma_n$
cancels with the one for the Gaussian random variables. Consequently, we
obtain (\ref{le}) because the mean of the Gaussian random variable $j_b$
is $x_b$.

The second identity (\ref{mq}) is obtained similarly to the first
one: It holds because the expectation $\<S_b\>$ changes to $\sigma_b\<S_b\>$
under the transformation for the Gaussian random variables and
$\<S_b\>^2$ does not change.

Now, to prove the first inequality (\ref{G1}), we consider the total
derivative with respect to $x_b$. The pressure function depends on $x_b$
through the Gaussian distribution and the nearest neighbor
potential. The derivative of the Gaussian distribution is rewritten as
follows:
\begin{eqnarray*} 
 \int \frac{dj_b}{\sqrt{2\pi}}\frac{d}{dx_b}
  \left(e^{-\frac{(j_b-x_b)^2}{2}}\right) f(j_b)
 &=& - \int \frac{dj_b}{\sqrt{2\pi}}\frac{d}{dj_b}
  \left(e^{-\frac{(j_b-x_b)^2}{2}}\right) f(j_b) \\
 &=& \left[\frac{d}{d j_b}f(j_b)\right],
\end{eqnarray*}
where we used integration by parts to obtain the final line. Thus we find
\begin{equation}
 \frac{dP}{dx_b} = \int \prod_{b'}\frac{dj_{b'}}{\sqrt{2\pi}}
 e^{-\frac{(j_{b'}-x_{b'})^2}{2}}\left(\frac{\partial}{\partial j_b}
 +\frac{\partial}{\partial x_b}\right) \ln \sum_{\{S\}}
 e^{\sum_{b}\beta_b j_b S_b}. \label{divx}
\end{equation}
Since a straightforward calculation yields
\begin{equation}
 \left(\frac{\partial}{\partial j_b}
 +\frac{\partial}{\partial x_b}\right) \ln \sum_{\{S\}}
 e^{\sum_{b}\beta_b j_b S_b} = x_b\<S_b\>+\<j_b S_b\>,
\end{equation}
we finally obtain from (\ref{le})
\begin{equation}
 \frac{dP}{dx_b} = x_b[\<S_b\>]+[\<j_b S_b\>]=x_b [\<S_b+1\>] \geq 0,
\end{equation}
which is the first inequality (\ref{G1}).

The proof of the second inequality (\ref{G2}) is a little more
complicated than that of the first one. The derivative by $x_b$ is
calculated similarly to (\ref{divx}) to obtain
\begin{equation}
 \frac{d}{dx_{b'}}[\<S_b\>] = [(x_{b'}+j_{b'})
 (\<S_b S_{b'}\>-\<S_b\>\<S_{b'}\>)].
\end{equation}
We can eliminate the Gaussian random variable $j_{b'}$ using the
following property of the Gaussian distribution:
\begin{equation}
  \left[(j_b-x_b) f(j_b)\right]
 = - \int \frac{dj_b}{\sqrt{2\pi}}\frac{d}{dj_b}
  \left(e^{-\frac{(j_b-x_b)^2}{2}}\right) f(j_b)
 = \left[ \frac{d}{d j_b}f(j_b)\right].
\end{equation}
Thus, we obtain
\begin{eqnarray*}
 \frac{d}{dx_{b'}}[\<S_b\>]&=&2x_{b'} [\<S_b S_{b'}\>-\<S_b\>\<S_{b'}\>]
  + \left[\frac{\partial}{\partial j_{b'}}
 (\<S_b S_{b'}\>-\<S_b\>\<S_{b'}\>)\right] \\
 &=& 2x_{b'}[\<S_b S_{b'}\>-\<S_b\>\<S_{b'}\>-
  \<S_b S_{b'}\>\<S_{b'}\>+\<S_b\>\<S_{b'}\>^2].
\end{eqnarray*}
Since the following set of identities can be proved similarly to
(\ref{mq}):
\begin{equation}
  [\<S_b S_{b'}\>] = [\<S_b S_{b'}\>^2]
\end{equation} 
\begin{equation}
 [\<S_b\>\<S_{b'}\>] = [\<S_b S_{b'}\>\<S_{b'}\>]
 =[\<S_b\>\<S_{b'}\>\<S_b S_{b'}\>]
\end{equation} 
\begin{equation}
 [\<S_b\>\<S_{b'}\>^2] = [\<S_b\>^2\<S_{b'}\>^2],
\end{equation}
we find
\begin{equation}
  \frac{d}{dx_{b'}}[\<S_b\>] = 2x_{b'}
   [(\<S_b S_{b'}\>-\<S_b\>\<S_{b'}\>)^2] \geq 0
\end{equation}
and the second inequality (\ref{G2}) has been proved.

\vskip 1truecm\par\noindent
{\bf Acknowledgments}.\\
One of us (P.C.) thanks Sandro Graffi for introducing him to the surface
pressure problems and the Tokyo Institute of Technology for the warm
hospitality and the stimulating atmosphere during the visit in which
this work was done. This work was supported by the Grant-in-Aid for
Scientific Research on Priority Area ``Statistical-Mechanical Approach
to Probabilistic Information Processing'' by the MEXT. P.C. was partially
supported by University of Bologna, Funds for Selected Research Topics and Funds
for Agreement with Foreign Universities.

\end{document}